\documentclass{emulateapj}
\usepackage{url}
\usepackage{gensymb}
\usepackage{textcomp}
\usepackage{xspace}
\usepackage{natbib,amsmath} \usepackage{graphicx}
\usepackage{hyperref} \usepackage{float} \usepackage{subfigure}
\usepackage{tabularx}
\providecommand{\e}[1]{\ensuremath{\times 10^{#1}}}
\newcommand{\um}{\textmu m\xspace}
\newcommand{\emcee}{\texttt{emcee}\xspace}
\newcommand{\nestle}{\texttt{nestle}\xspace}

\usepackage{array}
\newcolumntype{C}{>{$}c<{$}} 
\newcolumntype{R}{>{$}r<{$}} 
\usepackage{xcolor}

\begin{document}
\title{Forward modelling and retrievals with PLATON, a fast open source tool}

\author{ Michael Zhang\altaffilmark{1}, Yayaati Chachan\altaffilmark{2},
  Eliza M.-R. Kempton\altaffilmark{3,4}, Heather A. Knutson\altaffilmark{2} }

\altaffiltext{1}{Department of Astronomy, 
	California Institute of Technology, Pasadena, CA 91125, USA; 
	mzzhang2014@gmail.com}
\altaffiltext{2}{Division of Geological and Planetary Sciences,
  California Institute of Technology, Pasadena, CA 91125, USA}
\altaffiltext{3}{Department of Astronomy, University of Maryland, College Park, MD 20742, USA}
\altaffiltext{4}{Department of Physics, Grinnell College, 1116 8th Avenue, Grinnell, IA 50112, USA}

\begin{abstract}
  We introduce PLanetary Atmospheric Transmission for Observer Noobs (PLATON), a Python package that calculates transmission spectra
  for exoplanets and retrieves atmospheric characteristics based on observed
  spectra.  PLATON is easy to install and use, with common use cases taking no more than
  a few lines of code.  It is also fast, with the forward model taking much less
  than one second and a typical retrieval finishing in minutes on an ordinary desktop.  PLATON supports the most common atmospheric parameters, such as temperature, metallicity, C/O ratio, cloud-top pressure, and scattering slope.  It also has less commonly included features, such as a Mie scattering cloud model and unocculted starspot corrections.  The code is available online at https://github.com/ideasrule/platon under the open source
  GPL-3.0 license.
\end{abstract}


\section{Introduction}
\label{sec:introduction}
Transmission spectroscopy is an essential tool for understanding exoplanet
atmospheres.  By measuring the transit depth at different wavelengths,
astronomers have detected a variety of molecules, including many detections of water and a few of carbon monoxide \citep{madhusudhan_2014,deming_seager_2017}.  With sufficiently wide wavelength coverage, it is possible to constrain not
only compositions, but temperatures and cloud properties as well.  The inverse problem of retrieving atmospheric parameters given observed transit depths is crucial to the interpretation of observations.  The HST and Spitzer archives alone already hold observations of dozens of transiting exoplanets, and the volume of data will only increase as TESS discovers more planets around bright stars and ground-based transit spectroscopy matures.  A fast, easy to use, and open source retrieval code is needed in order to help the community quickly extract atmospheric parameters from these observations.

The exoplanet community has developed many retrieval codes over the years, as reviewed in \cite{madhusudhan_2018}.  The first retrieval code was published in \cite{madhusudhan_2009}, but many more have been developed since then, including \cite{benneke_2012}, \cite{line_2013}, and \cite{waldmann_2015}.  These codes use a wide variety of techniques.  Most use either MCMC or nested sampling to map the posterior distribution, but some use self-consistent temperature-pressure profiles while others use parametric profiles; some enforce chemical equilibrium while others allow individual molecular abundances to vary freely; some compute eclipse depths while others only compute transit depths; some use opacity sampling while others use the correlated-k method.  Despite the large number of codes, we were only able to find one documented, publicly available retrieval code with a dedicated paper \citep[Tau-REX;][]{waldmann_2015}. The Bayesian Atmospheric Radiative Transfer (BART) code is partially described in one subsection of a dissertation \citep{blecic_2016b} and one subsection of a paper \citep{blecic_2017}.  CHIMERA \citep{line_2013} appears to be publicly available\footnote{\url{https://github.com/ExoCTK/chimera}}, but has no documentation or examples.

Although the papers describing these codes rarely have benchmarks, it appears that a typical retrieval takes days to weeks on a typical desktop computer with 4 cores \citep{zingales_2018}.  We aimed to create a publicly available fast retrieval code that can be run on a typical desktop, and takes minutes to hours instead of days to weeks.

Recently, \cite{kempton_2017} introduced Exo-Transmit.  Exo-Transmit calculates transmission spectra for exoplanets of arbitrary size and composition, with pre-packaged data files that make it easy to specify common compositions and pressure-temperature profiles.  However, with a run time on the order of tens of seconds, Exo-Transmit is too slow to incorporate into a retrieval code.  Exo-Transmit also relies on pre-packaged data files to specify the composition of the atmosphere, and does not allow the user to specify arbitrary metallicities, C/O ratios, or abundances.  In addition, Exo-Transmit is written in C and therefore requires additional steps to interface with Python-based codes.

Inspired by Exo-Transmit, we developed PLATON to address these issues.  PLATON was written from scratch in pure Python, and supports both Python 2.7 and 3.5 on both Linux and Mac machines.  It uses the same opacity data and a few of the same algorithms as Exo-Transmit.  Its forward model contains all the functionality of Exo-Transmit, but is 100-1000x faster--fast enough to incorporate into a Markov chain Monte Carlo (MCMC) or nested sampling retrieval.  These speed improvements come from three main sources.  First, it loads all data files once and holds them in memory afterwards.  Second, it performs transit depth calculations only for the wavelength ranges requested by the user, and only for atmospheric layers above the cloud pressure.  Third, the radiative transfer is formulated as a matrix-vector multiplication, which \texttt{numpy} can compute very efficiently by calling a highly optimized Basic Linear Algebra Subsystems (BLAS) routine.  This decreases the computation time from 1-2 seconds to 30 milliseconds.  Aside from these optimizations, PLATON also incorporates several new features, including the ability to model Mie scattering, account for unocculted starspots, compute equilibrium abundances at user-specified metallicities and C/O ratios, and account for varying gravitational acceleration as a function of height for extended atmospheres.  This forward modelling code was then incorporated into a retrieval module within PLATON.

PLATON is in many ways complementary to Tau-REX.  Whereas Tau-REX considers a small number of molecules (which could be automatically chosen by a machine learning algorithm), PLATON considers a large number of pre-determined molecules.  Whereas Tau-REX performs free retrievals, allowing for the abundances of individual molecules to be retrieved, PLATON assumes equilibrium chemistry throughout the atmosphere.  In ease of installation and use, however, PLATON has a clear advantage.  Installing Tau-REX is a complex multi-step process which involves signing up for an account, installing CMake, OpenMPI, and MultiNest manually, soft linking libraries (on Macs), installing some Python packages, compiling Tau-REX's C++ and Fortran libraries, and downloading a separate set of input data.  After installation, the first author ran into numerous problems running the tests, mostly related to problems with the dependencies.  In contrast, PLATON's installation process is a single line, which also installs all dependencies:

\texttt{pip install platon}
\newline Or alternatively, after cloning the Git repository:

\texttt{python setup.py install}
\newline We describe PLATON's underlying model structure and validation in Section 2, and its incorporation into a retrieval framework in Section 3.

\section{Forward model}
\label{sec:model}
Transmission spectroscopy has the benefit of being relatively simple to
model.  By its nature, it probes higher altitudes (P $<$ 1 bar), where the atmosphere can be approximated as isothermal \citep{burrows_2008}.  In addition, although nightside pollution of transit depths will be significant for JWST \citep{kipping_2010}, planetary emission is negligible compared to transmission of starlight.  This drastically simplifies the radiative transfer calculation and limits the amount of model uncertainty.  Our forward model uses the opacity files from Exo-Transmit, in addition to some of its algorithms.  This section will discuss the algorithms and opacities briefly, but more detail can be found in \cite{kempton_2017}.

\subsection{Overview}
To calculate a transmission spectrum, we assume an isothermal atmosphere with
equilibrium chemistry.\footnote{Arbitrary P-T profiles and compositions are also supported, but this is not the recommended use case.  We refer the user to the documentation for information on these features.}
We first divide the atmosphere into 500 layers, equally spaced in log P from $10^{-4}$ to $10^8$ Pa. The composition of each layer is calculated assuming equilibrium chemistry, from which follows the mean molecular weight.  The physical depth of each layer is then obtained by solving the hydrostatic equation:

\begin{equation}
\frac{dP}{dr} = -\frac{GM}{r^2} \frac{\mu m_{amu}P}{kT}
\end{equation}

The integration constant is set by P(R) = 1 bar: that is, the radius which the user specifies for the planet is taken to be the radius at a reference pressure of 1 bar.  The reference pressure is arbitrary, but we picked 1 bar because it is easy to remember and close to standard atmospheric pressure on Earth.  After solving the hydrostatic equation, the opacity of each layer is then calculated from its composition.  We include gas absorption, collisional absorption, and Rayleigh scattering, the last of which can be modified with a user-specified slope and amplitude.

Finally, we carry out the radiative transfer calculation. Consider a ray passing through the
atmosphere with impact parameter $r$, where $r$ is the distance from planet center to one of the layers.  This ray passes through all layers of the atmosphere above $r$, but lingers in each layer for different distances.  By calculating the distance it takes to traverse each layer and multiplying by that layer's absorption coefficient, we calculate the optical depth experienced by this ray:

\begin{equation}
\tau_{\lambda}(r_i) = \sum_{j=i}^N \alpha_{\lambda}(j) \Delta l(i, j)
\end{equation}
The transit depth can then be calculated by adding up the cross sectional areas of each layer, weighted by the amount of light let through:

\begin{equation}
D_{\lambda} = (R_{bot}/R_s)^2 + 2\sum_{i=1}^N \frac{r dr}{R_s^2} (1-e^{-\tau_{\lambda}})
\end{equation}
where $R_{bot}$ is the radius at the bottom-most layer.

As a final step, we correct the transit depth for unocculted star spots.  This step is necessary for active stars because the hot unspotted regions have a different spectrum from the colder star spots, and a planet that blocks only the former will appear bigger than it actually is.  Worse, this radius inflation depends on wavelength, giving rise to a spurious transit spectrum shape \citep{mccullough_2014, rackham_2018}.  To correct for this effect, we use:

\begin{equation}
D_{\lambda, c} = D_{\lambda} \frac{S(\lambda, T_{clear})}{fS(\lambda, T_{spot}) + (1-f)S(\lambda, T_{clear})}
\end{equation}
where $f$ is the spot fraction, $T_{spot}$ is the spot temperature, $T_{clear}$ is the temperature of unspotted regions, and $S$ is the spectrum of the stellar surface.  We obtain both $S(\lambda, T_{clear})$ and $S(\lambda, T_{spot})$ by interpolating the BT-NextGen (AGSS2009) stellar spectral grid \citep{allard_2012}, as provided by the Spanish
Virtual Observatory\footnote{\url{http://svo2.cab.inta-csic.es/theory/newov2/index.php}}.  We neglect any contributions to the spot spectrum that are not purely due to temperature.

\subsection{Absorption coefficients}
We use the same data files as Exo-Transmit for gas absorption and collisional absorption coefficients, although these will be updated in version 3.  Per-species gas absorption coefficients were computed on a wavelength-temperature-pressure grid using line lists, with a resolution of $\lambda/\Delta \lambda = 1000$ over the range 0.3-30 $\mu m$.  The sources of these line lists are shown in Table 2 of \cite{luppu_2014}.  For most molecules, the line list comes from HITRAN, but they use miscellaneous other sources, such as \cite{freedman_2008, freedman_2014}.  All collisional absorption coefficients are from HITRAN.  We refer the reader to \cite{kempton_2017} for more information.

For Rayleigh scattering, we compute the absorption coefficient using the equation:

\begin{equation}
\alpha_\lambda = \frac{128}{3\pi^5}\frac{P}{kT}p^2 \lambda^{-4}
\end{equation}
where $p$ is the polarizability of a species.  For both Exo-Transmit and PLATON, polarizabilities are obtained from the CRC Handbook of Chemistry and Physics.  We use an updated list of polarizability values, including several species assumed to have zero polarizability in Exo-Transmit: H, O, C, N, C\textsubscript{2}H\textsubscript{4}, H\textsubscript{2}CO, OCS, OH, SO\textsubscript{2}, Na, and K.  The polarizability of OH is not in the CRC handbook, and was instead taken from \cite{pluta_1988}.  A few species have no published polarizability values that we can find: MgH, SH, SiH, SiO, TiO, and VO.  We adopt 0 for their polarizability.

We compute total absorption coefficients at temperature and pressure grid points bordering the atmospheric temperature-pressure profile.  2D bilinear interpolation is then used to get the coefficients at each point along the temperature profile.  This is the same algorithm used by \cite{kempton_2017}.

\subsection{Atmospheric composition}
In the sphere of atmospheric composition, we depart from Exo-Transmit by providing an equilibrium chemistry model computed with \texttt{GGchem} \citep{woitke_2018}.  The user can provide two parameters: metallicity and C/O ratio.  If the user does not specify these parameters, we assume solar metallicity and solar C/O ratio (0.53).  From these parameters, we compute the abundances of 34 atomic and molecular species in every layer of the atmosphere.  These 34 species are the same as those included in \cite{kempton_2017}, except that we leave out C\textsubscript{2}H\textsubscript{6} and SH because these species are not included in the default \texttt{GGchem} databases.

While many public codes exist for equilibrium chemistry \citep{blecic_2016a, woitke_2018, stock_2018}, they are currently too slow for use in retrievals.  This is why we opted to use \texttt{GGchem} to generate a grid of abundances.  \texttt{GGchem} computes equilibrium species abundances from atomic abundances by minimizing the total Gibbs free energy of the mixture.  We generate separate grids, one with and the other without condensation.

The abundance grids have 5 dimensions: species name, temperature, pressure, metallicity, and C/O ratio.  The temperature and pressure grid points are chosen to match the opacity grid, with temperature ranging from 300 to 3000 K in 100 K intervals and pressure from $10^{-4}$ to $10^{8}$ Pa in decade intervals.  The pressure range is determined by our opacity data, which are only available for this range.  The temperature range is set mostly by our opacity data, which span 100-3000 K at 100 K intervals.  We did not generate abundances down to 100 K because \texttt{GGchem} becomes unstable at very low temperatures.  Metallicity ranges from $\log_{10}(Z)=-1$ to $\log_{10}(Z)=3$ in steps of 0.05, while C/O ratio ranges from 0.2 to 2.0 in steps of 0.2.  We then perform 4D linear interpolation over the grid, obtaining log(Abundance) as a function of T, log(P), log(Z), and log(C/O).  The relative error introduced by interpolation in metallicity is on the order of $10^{-4}$.  This is significantly smaller than the error introduced by interpolation in C/O, which is on the order of a few percent.   Some molecules (such as water and HCN) have a sharp abundance transition at C/O $\sim$ 1, giving rise to interpolation errors of tens of percent in this regime.  However, the abundances change so rapidly around C/O $\sim$ 1 that our assumption of uniform equilibrium abundances across the whole planet also becomes invalid, as it is unlikely that the entire planet has exactly the same C/O ratio.  This makes the interpolation errors less significant.  Nevertheless, our development branch contains an abundance grid with twice the resolution in C/O, increasing to 4x the resolution around C/O $\sim$ 1.  Interpolation errors are less than 1\% in almost all cases with this new grid.  The new grid will be incorporated in our next release.

For retrievals, the current data is almost never good enough to constrain any parameter to better than one grid spacing.  For forward models, we recommend using parameters that correspond exactly to a grid point in temperature, metallicity, and C/O to avoid interpolation errors.

\subsection{Clouds and hazes: parametric}
By default, PLATON accounts for clouds and hazes by allowing the cloud-top pressure, scattering amplitude, and scattering slope to vary as free parameters in the fit.  The cloud-top pressure defines the height in the atmosphere below which no light can penetrate.  The user specifies this parameter, or can set it to infinity for a clear atmosphere.  The scattering amplitude and slope are a simple way to parameterize scattering properties without invoking a microphysical model.  The default value--an amplitude ($A$) of 1 and a slope of 4--corresponds to pure Rayleigh scattering.  If $A$ is changed while the slope is fixed at 4, the absorption coefficient for scattering is simply multiplied by $A$ at all wavelengths.  If the slope $s$ is not 4, we set the scattering absorption coefficient such that it is $A$ times the Rayleigh absorption at the reference wavelength of 1 $\mu m$, and is proportional to $\lambda^{-s}$ at all wavelengths.  This parameterization makes no assumptions about the underlying physics while allowing users to see at a glance how strong the scattering is compared to Rayleigh at the reference wavelength (by comparing $A$ to 1), and how the scattering strength behaves with wavelength (by looking at the slope).  Since the reference wavelength is only a matter of definition and does not change the underlying model, the user can set it to any value.

\subsection{Clouds and hazes: Mie scattering}
In addition to the parametric approach, PLATON supports Mie scattering (Benneke et al., subm.).  Clouds and hazes often have particles in the micron range, comparable to the wavelengths of most transit spectra observations.  This makes Mie scattering an important component of atmospheric physics \citep{marley_2013}.  For example, Benneke et al., subm. use Mie scattering to explain the anomalously low 3.6 and 4.5 \um Spitzer transit depths for GJ 3470b as compared to its WFC3 (1.1--1.6 \um) transit depths.

In the Mie scattering mode, the user specifies a cloud-top pressure ($P_{cloud}$), a complex refractive index ($m = n - ik$), a mean particle size ($r_m$), a geometric standard deviation for particle size ($\sigma_g$), a maximum number density ($n_0$), and a fractional scale height ($f$).  The atmosphere is assumed to contain particles of the same refractive index at all altitudes above the ``cloud top pressure".  Below this pressure, the atmosphere is assumed to be perfectly opaque, corresponding to perfectly opaque clouds.  The number density is given by:

\begin{equation}
n = n_0 e^{-\frac{h}{fH_{gas}}},
\end{equation}

where h is the height above the cloud top, f is specified by the user, and $H_{gas}$ is the gas scale height. 

The extinction cross section of a single particle of radius r is given by:

\begin{equation}
\sigma = \pi r^2 Q_{ext}(m, 2 \pi r/\lambda)
\end{equation}
To calculate $Q_{ext}$, we use the same algorithm as LX-MIE \citep{kitzmann_2018}, except implemented in Python instead of C++.  This algorithm is fast, simple, stable, and does not lead to overflows.  Since the calculation of $Q_{ext}$ is time intensive (often taking hundreds of milliseconds), we cache the results of every calculation for the lifetime of the transit depth calculator object.  Every time the value of $Q_{ext}(m, x)$ is required, the cache is first consulted.  If at least one value in the cache has the same $m$ and an $x$ within 5\% of the requested $x$, we perform linear interpolation on all cache values with the same $m$ and return the interpolated value.  If no cache value satisfies these criteria, we consider this a cache miss.  $Q_{ext}$ is then calculated for all cache misses and added to the cache.

We assume that the particles follow a log-normal radius distribution:

\begin{equation}
P(r) = \frac{1}{\sigma r \sqrt{2 \pi}} \exp\Big(-\frac{(\ln{r} - \ln{r_m})^2}{2\sigma_g^2}\Big),
\label{eq:lognormal_dist}
\end{equation}
in which case the effective cross section of one particle becomes:

\begin{equation}
\bar{\sigma}(\lambda) = \int_0^{\infty} \pi r^2 Q_{ext}(m, \frac{2 \pi r}{\lambda}) \frac{1}{\sigma \sqrt{2 \pi} r} e^{-\frac{(\ln{r} - \ln{r_m})^2}{2\sigma_g^2}} dr
\end{equation}
After a change of variables to $z = \frac{\ln{r} - \ln{r_m}}{\sigma}$, we have:

\begin{equation}
\bar{\sigma}(\lambda) = \int_{-\infty}^{\infty} e^{-z^2/2} \frac{\sqrt{\pi}}{2} r_m^2 e^{2\sigma z}Q_{ext}(m, \frac{2 \pi r_m e^{\sigma z}}{\lambda}) dz
\end{equation}
We integrate this equation by computing the integrand at 100 different values of z ranging from -5 to 5, then using the trapezoid rule.  These values were chosen to be densely spaced near 0 and less densely spaced at high or low z.  The number of points, the range of z values, and the spacing of z values were chosen so that the integral is as accurate as possible over a wide range of refractive indicies and mean particle sizes without being too computationally expensive.

\subsection{Wavelength binning}
The user can request the transit depth in specific wavelength bins.  In these situations, PLATON uses the stellar spectrum to compute a properly binned transit depth.  This is important for broad-band instruments like Spitzer/IRAC, where the stellar spectrum changes significantly from one side of the bandpass to the other.

Using the user specified stellar temperature, PLATON interpolates the BT-NextGen (AGSS2009) stellar spectral grid to get a stellar spectrum, assuming log(g) = 4.5 (typical for main sequence stars) and solar metallicity.  If the user also specifies a spot coverage fraction and spot temperature, PLATON computes the spot spectrum by interpolation, then computes the weighted average of the unspotted and spotted spectrum.  The spectrum, which is in units of erg/cm\textsuperscript{2}/s/A, is converted to a photon flux and multiplied by the spacing between adjacent wavelengths in our wavelength grid.  The multiplication is necessary because our wavelength grid is uniform in log space, but not in linear space, as each grid spacing is 0.1\% larger than the previous.  Finally, the unbinned transit depths are converted to a binned transit depth by a weighted average, the weight at each wavelength being the photon flux multiplied by the grid spacing.

\subsection{Validation}

\begin{figure}[ht]
  \centering 
  \includegraphics[width=\linewidth]{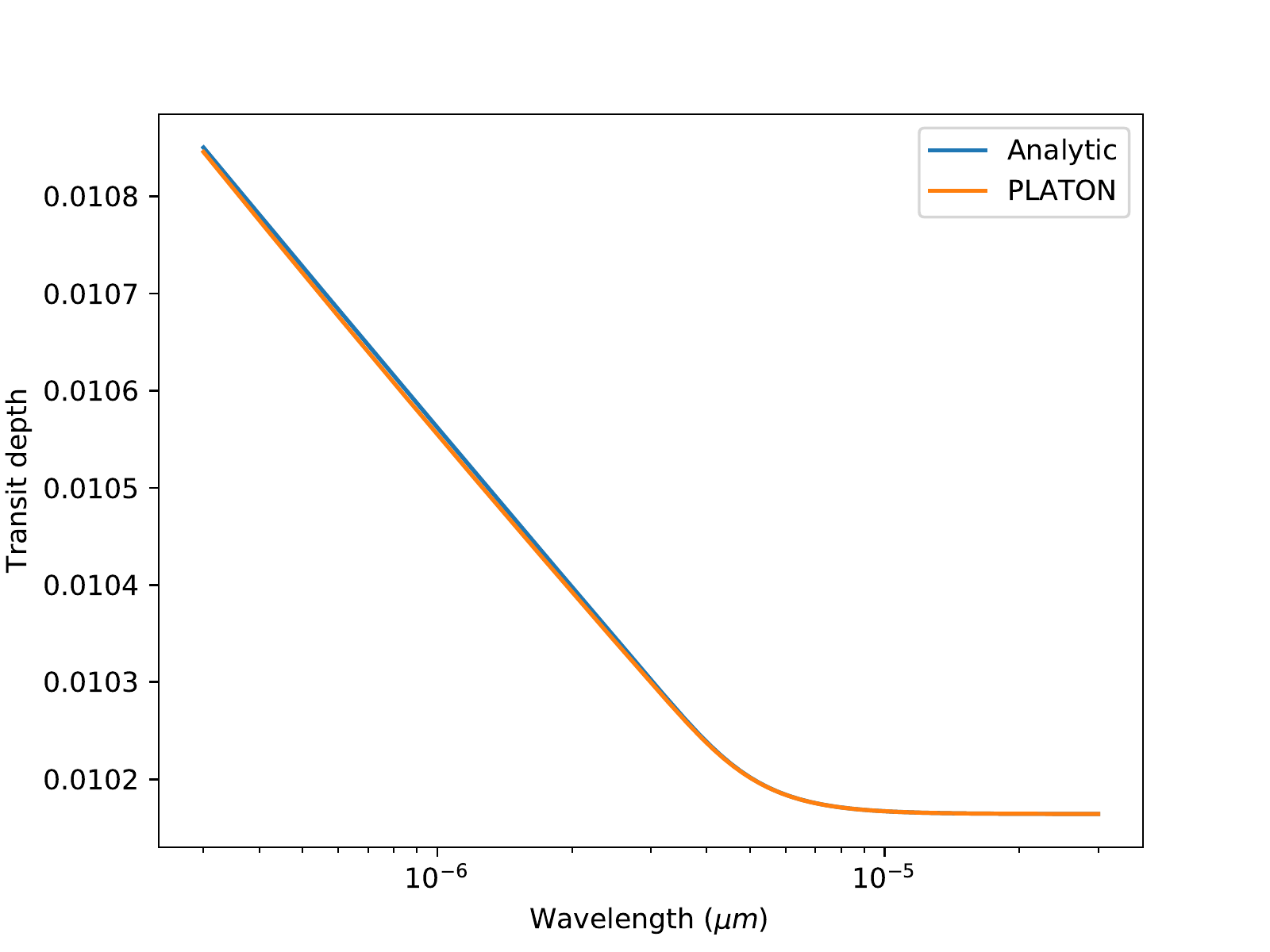}
  \caption{Transit depths with all opacity except Rayleigh scattering turned off.  The transit depth obeys a power law to within 0.5\% blueward of 3 $\mu m$, as derived by \cite{shabram_2011}.  Redward of 3 $\mu m$, transit depths flatten out because the atmosphere becomes so transparent that the photospheric pressure becomes greater than $10^8$ Pa, the maximum pressure we consider.}
\label{fig:power_law_haze}
\end{figure}

We validate our code with unit tests covering every component of the package.  In cases where the expected answer can be computed analytically with simplified inputs, such as the interpolation routines or the radiative transfer, we check against these analytic results. An example is shown in Figure \ref{fig:power_law_haze}, where we reproduce the theoretical transit depths for a hydrogen atmosphere with only Rayleigh scattering.  We make use of Equation 27 of \cite{betremieux_2017} to derive:

\begin{align}
    \kappa &= \frac{128\pi^5}{3} \frac{p^2}{\lambda^4 \mu m_H}\\
    \tau_s &= \frac{P_s\kappa}{g} \sqrt{\frac{2\pi R_s}{H}}\\
    R &= R_s + H (\gamma + \ln(\tau_s) + E_1(\tau_s)),
\end{align}

where H is the scale height, $\gamma = 0.57721$ is a constant, p=0.8059\e{-24} cm\textsuperscript{3} is the polarizability of the hydrogen molecule, $R_s$ is the radius at the surface, and $\tau_s$ is the slant optical depth at the surface, and $E_1$ is the exponential integral.  PLATON does not model a real surface, but we do not model radiative transfer at pressures greater than $10^8$ Pa, creating an effective surface at that pressure.

In cases where it is possible to compare against published results, such as the $Q_{ext}(m,x)$ calculations for selected $m$ and $x$, we verify that we reproduce the published results in \cite{kitzmann_2018}.  For methods where such a check would be impractical or take too long, such as the retrieval, we simply ensure the method runs without error and returns a result in the expected format.  We use Travis CI to ensure that all unit tests are run after every commit.  Travis CI creates pristine virtual machines after every commit, installs PLATON on each machine, and runs the unit tests to make sure they pass.  We use four virtual machines to test PLATON in four configurations: Ubuntu with Python 2.7, Ubuntu with Python 3, OS X with Python 2.7, and OS X with Python 3.  This way, we ensure that PLATON can be installed and successfully run on the most common configurations.

\begin{figure}[ht]
  \centering \subfigure[Full wavelength coverage] {\includegraphics
    [width=0.5\textwidth]{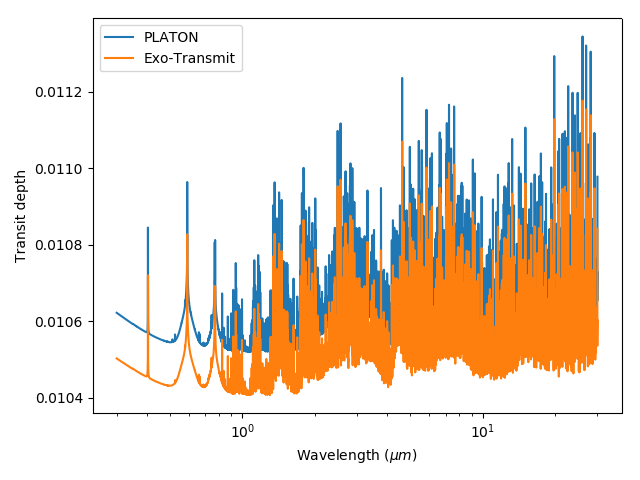}}\qquad
  \subfigure[Zoomed into WFC3 band] {\includegraphics
    [width=0.5\textwidth]{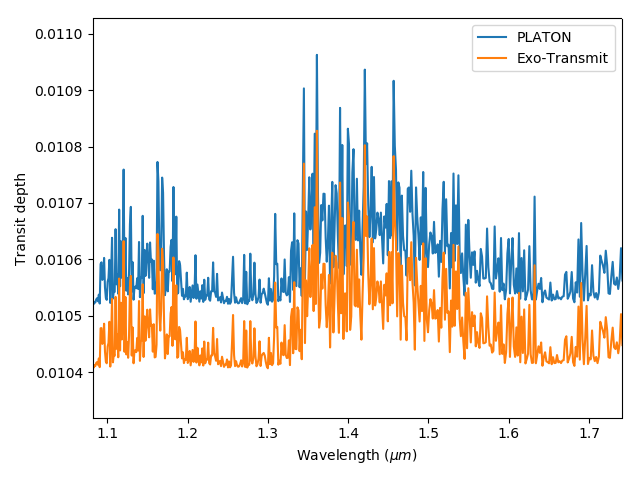}}
    \caption{Transmission spectrum of a 1200 K Jupiter mass, Jupiter radius planet orbiting a Sun-like star, as computed by PLATON and Exo-Transmit.  The difference is mainly due to PLATON truncating the atmosphere at a much lower pressure.}
\label{fig:platon_exotransmit}
\end{figure}

\begin{figure}[ht]
  \centering \subfigure[Full wavelength coverage] {\includegraphics
    [width=0.5\textwidth]{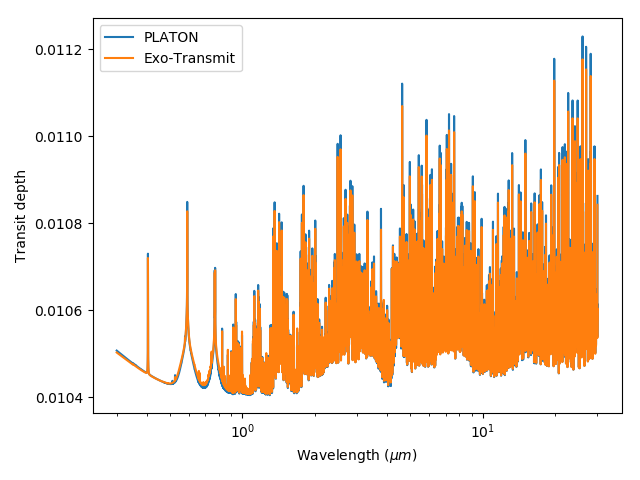}}\qquad
  \subfigure[Zoomed to show individual lines] {\includegraphics
    [width=0.5\textwidth]{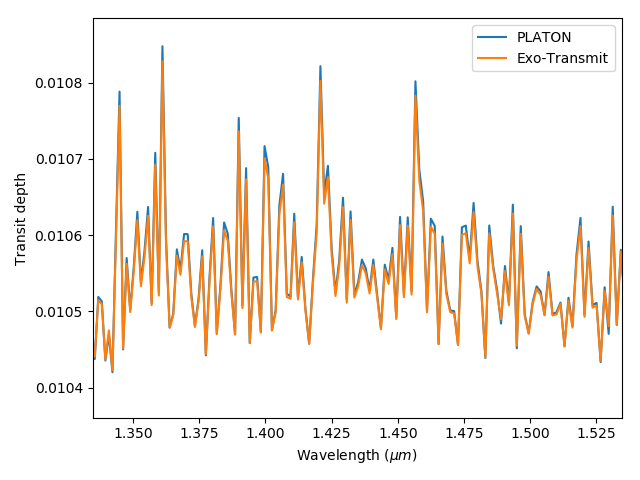}}
    \caption{Same as Figure \ref{fig:platon_exotransmit}, but with PLATON transit depths shifted downwards by 115 ppm.}
\label{fig:platon_exotransmit_offset}
\end{figure}

We also compare our transmission spectrum to the output of Exo-Transmit.  In an old version of the code, which was intended to match Exo-Transmit output exactly, the transmission spectrum matched to within machine precision for all wavelengths.  In the current version, we have made many design decisions that increase accuracy or usability at the cost of breaking the exact match between the two codes.  Figure \ref{fig:platon_exotransmit} shows a comparison between PLATON and Exo-Transmit spectra for a typical hot Jupiter.  The main difference between the two is a uniform shift, with PLATON reporting 115 ppm deeper transit depths.  Figure \ref{fig:platon_exotransmit_offset} shows that if this shift is removed, the two spectra agree almost perfectly.

There are two reasons why PLATON reports higher transit depths.  First, PLATON correctly takes into account the decline of gravity with height, while Exo-Transmit assumes constant gravity.  This makes PLATON's atmospheres larger--an effect that is most pronounced for super-Earths, but discernible even for Jupiters.  Second, Exo-Transmit truncates the atmosphere, taking into account only the region between 0.1 Pa and $10^5$ Pa.  We consider the entire pressure range for which we have absorption data, spanning $10^{-4}$ Pa to $10^8$ Pa.  In practice we have found that increasing the upper pressure limit has no effect because the photosphere is well above $10^5$ Pa, but decreasing the lower pressure limit does increase the transit depth substantially.  This is especially the case at high metallicities, where strong lines saturate and the transit depths of strong lines correspond to the 0.1 Pa height limit in Exo-Transmit.

\begin{figure*}[ht]
  \centering \subfigure[Full wavelength resolution] {\includegraphics
    [width=0.45\textwidth]{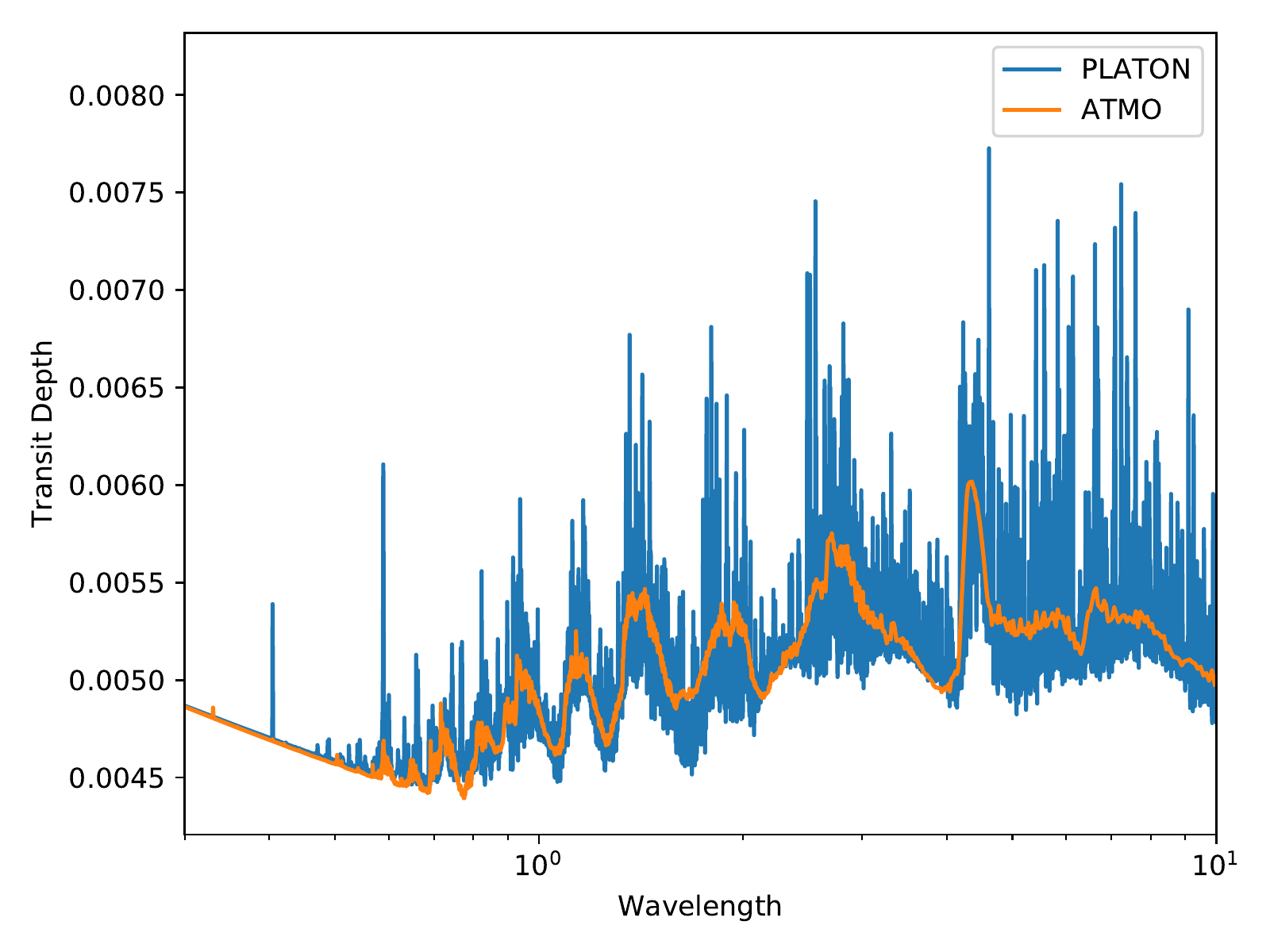}}\qquad
  \subfigure[Rough HST resolution] {\includegraphics
    [width=0.45\textwidth]{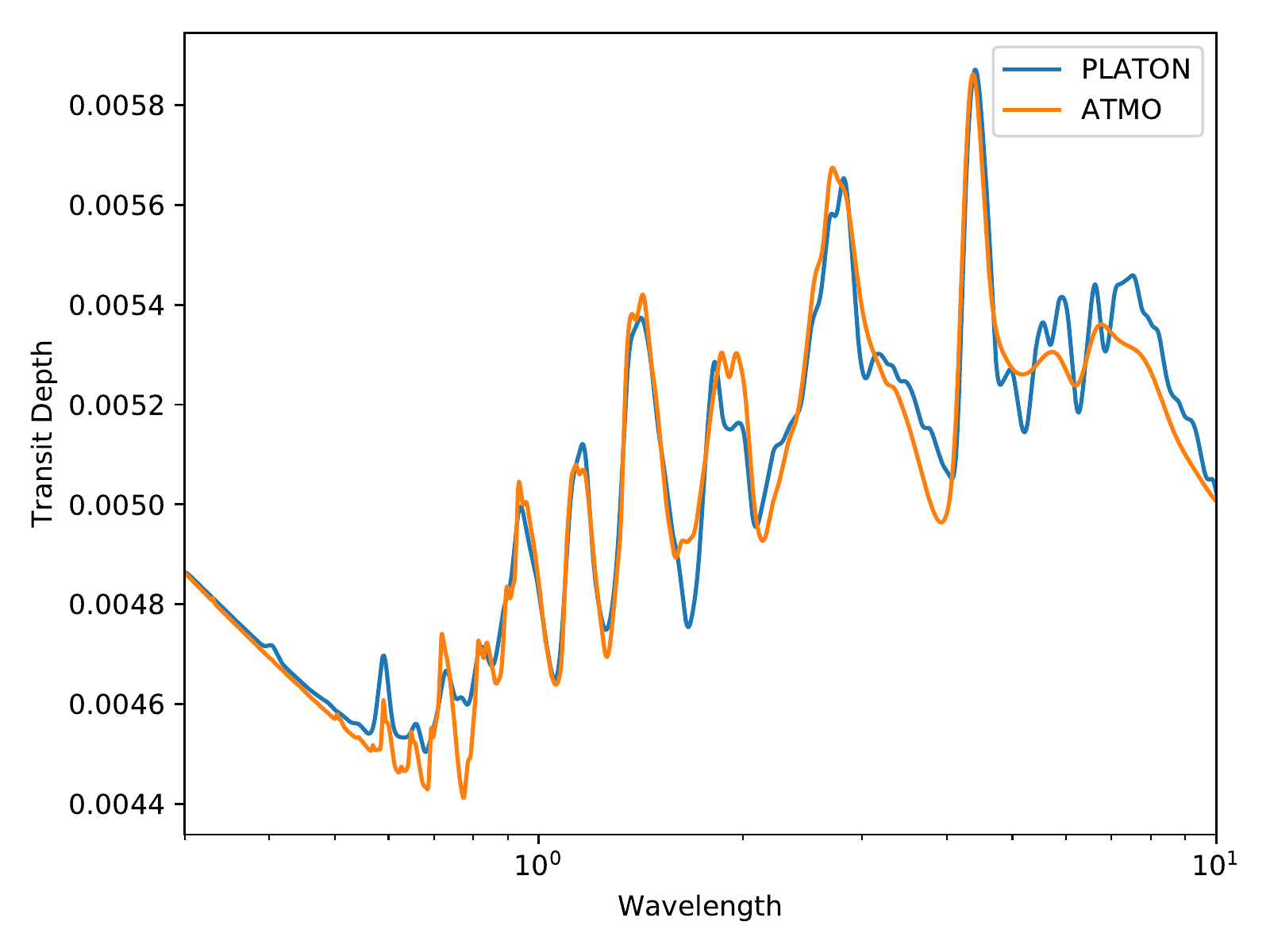}}
    \caption{Transmission spectrum for HAT-P-26b generated from ATMO \citep[obtained from][]{goyal_2018a} and PLATON for an isothermal atmosphere with temperature 851 K, 10x solar metallicity, C/O ratio of 0.35, and haze factor/scattering factor of 1. In the right panel, we show the transmission spectrum smoothed by a Gaussian filter of $\sigma = 15$ to roughly match the resolution of HST.}
\label{fig:platon_atmo}
\end{figure*}

As a final validation step, we compare our forward model for HAT-P-26b against that of another atmospheric code, ATMO \citep{goyal_2018b}.  The result is shown in Figure \ref{fig:platon_atmo}.  Due to our opacity sampling algorithm, our transit depths are much spikier than ATMO's, but the transit depths match decently once they are binned to the resolution of HST.

\section{Retrieval}
We support retrievals of atmospheric parameters with either MCMC or multimodal nested sampling.  MCMC is implemented by \emcee \citep{foreman-mackey_2013}, which uses an affine invariant ensemble sampler.  Multimodal nested sampling was first described by \cite{shaw_2007}, but we use the pure Python implementation \nestle, available on Github at https://github.com/kbarbary/nestle.  With the exception of the complex refractive index, any parameter that can be passed to the forward model can be included in the retrieval.  These parameters are shown in Table \ref{table:parameters}.  The last parameter in Table \ref{table:parameters}, error multiple, multiplies every observational error.  It accounts for over- or under- estimation of error bars by the user, assuming all error bars are off by the same factor.

\begin{table*}[t]
  \centering
  \begin{tabular}{c c c c}
  \hline
  	Parameter name & Supported range & Log space? & Default\\
      \hline
      Stellar radius & Any & No & None\\
	  Stellar temperature & Any & No & None\\
	  Stellar spot fraction & Any & No & 0\\
      Stellar spot temperature & Any & No & None\\
      Planet mass & Any & No & None\\
      Planet radius (at 1 bar) & Any & No & None\\
      Temperature & 300-3000 K & No & None\\
      Metallicity & 0.1-1000x solar & Yes & Solar\\
      C/O ratio & 0.2-2.0 & No & 0.53\\
      Cloud-top pressure & $\infty$, or $10^{-4}$ - $10^8$ Pa & Yes & $\infty$\\
      Scattering amplitude & Any & Yes & 1\\
      Scattering slope & Any & No & 4\\  
      Error multiple   & Any & No & 1\\
      Particle number density & Any & Yes & 1 $m^{-3}$\\
      Particle radius & Any & Yes & 1 $\mu m$\\
      Particle radius standard deviation & Any & No & 0.5\\
      Particle fractional scale height & Any & No & 1\\
  \end{tabular}
          \caption{Parameters that can be retrieved.  PLATON supports a limited range for some parameters, as shown in the second column.  The third column indicates whether PLATON takes the base-10 log of the parameter during retrieval.}
  \label{table:parameters}
\end{table*}

The user can choose to fit or freeze any parameter.  However, if the user chooses to use the parametric method to account for clouds and hazes, the last 4 parameters in Table \ref{table:parameters} are irrelevant. PLATON requires that the log particle number density be set to $-\infty$ to avoid confusion.  If the user chooses to use Mie scattering, the complex refractive index must be specified, along with the last four parameters in the table.  PLATON requires that the log scattering amplitude be set to the default value of 0 and the slope be set to the default value of 4, again to avoid confusion.  If a parameter is included in the fit, the user can choose a uniform prior or a Gaussian prior.  Gaussian priors are most appropriate for parameters like the planet mass, where a published value exists, but with error bars substantial enough to broaden the distribution of the atmospheric parameters.

\subsection{Benchmarks}

We benchmark PLATON on a typical desktop computer to illustrate its performance.  The computer runs Ubuntu 16.04 LTS with a Core i7 7700k CPU and 16 GB of RAM.  PLATON never uses more than 500 MB of memory, since that is the total size of its data files, and should perform just as well on computers with less than 16 GB of RAM.

\begin{table}[ht]
  \centering
  \caption{Benchmarks for desktop computation of forward model}
  \begin{tabular}{c c c}
  \hline
  	Band & Wavelength range ($\mu m$) & Time (ms)\\
   \hline
    All wavelengths & 0.3-30 & 393\\
    STIS & 0.293-1.019 & 57\\
    WFC3 & 1.119-1.628 & 29\\
    Spitzer 3.6 $\mu m$ & 3.2-4.0 & 16\\
    Spitzer 4.5 $\mu m$ & 4.0-5.0 & 16\\
      \hline
  \end{tabular}
  \label{table:forward_model_benchmarks}
\end{table}

When the forward model is first initialized, PLATON loads all relevant data files into memory.  This takes 390 ms, but is only done once.  Table \ref{table:forward_model_benchmarks} shows the amount of time taken to compute transit depths within the most commonly used bands once PLATON is initialized.  The time taken most directly depends on the number of wavelength grid points within the band.  Since grid points are spaced uniformly in logarithmic space with R=1000, the number of grid points is proportional to the ratio between the maximum and minimum wavelengths.  In the future, if we increase the resolution we support, run times will increase proportionally.  It should be noted that all values in the table are computed for HD 209458b, assuming a clear atmosphere and no Mie scattering.  If there are clouds, PLATON neglects the part of the atmosphere below the cloud-top pressure, substantially improving its performance.  If Mie scattering is turned on, the calculation of $Q_{ext}$ can consume hundreds of milliseconds on the first run of the forward model.  However, every run populates the $Q_{ext}$ cache, and a future run with similar parameters will have a high cache hit rate. Mie calculations will then consume negligible time.

\begin{table*}[ht]
  \centering
  \caption{Benchmarks for retrieval using real HD 209458b data}
  \begin{tabular}{c c c c}
  \hline
  	Band(s) & Algorithm & Time (min) & Likelihood evaluations\\
   \hline
    WFC3 & nested sampling & 6.3 & 20,999\\
    STIS + WFC3 + Spitzer & nested sampling & 29 & 17,205\\
    WFC3 & MCMC & 13 & 50,051\\
    STIS + WFC3 + Spitzer & MCMC & 57 & 50,051\\
     \hline
  \end{tabular}
  \label{table:retrieval_benchmarks}
\end{table*}
 
Table \ref{table:retrieval_benchmarks} shows the typical performance of retrievals.  Users can run the second and fourth benchmarks themselves by running \texttt{examples/retrieve\_multinest.py} and \texttt{examples/retrieve\_emcee.py}, respectively.  These benchmarks use published HD 209458b data from \cite{knutson_2007}, \cite{deming_2013}, and \cite{evans_2015}.  The nested sampling runs were performed with 100 live points, and the MCMC runs had 50 walkers and 1000 steps.  These correspond to 17,000 and 50,000 likelihood evaluations respectively.  The number of likelihood evaluations, and hence the run time, is proportional to the number of live points or steps.

For MCMC, we verified convergence in two ways.  First, we ran another retrieval with 10,000 steps and checked that it gave similar posteriors.  Second, we estimated the autocorrelation length using a variant of the ``new'' method recommended by the author of emcee, Dan Foreman-Mackey, on his webpage\footnote{\url{https://emcee.readthedocs.io/en/latest/tutorials/autocorr}}.  Namely, for each dimension corresponding to a parameter, we compute the autocorrelation length of the chain, flattened along the walker dimension.  The estimates are averaged together to obtain an average autocorrelation length.  If the chain is less than 50 times longer than this average, we consider the estimate unreliable and run more iterations.  If the chain is more than 50 times longer, we consider the estimate reliable, and the chain to have converged.  The factor of 50 comes from the recommendation of the same webpage.  For our benchmark retrieval, we find an autocorrelation length of 7 steps--corresponding to 350 samples, as we have 50 walkers.  For the chain with 10,000 steps, we compute an autocorrelation length of 24 steps.  Both numbers are far smaller than the total number of steps, but the fact that the estimates are discrepant shows the disturbing fact that the autocorrelation length estimate increases with the number of samples.  This is a property of the algorithm, not of our chains.  It is one of the reasons why monitoring convergence is not trivial, and is why we have decided to refer the user to the extensive online and published literature on this topic.

\subsection{Comparison with retrieval results in literature}
To evaluate the performance of our retrieval tool, we focus on planets with large datasets and published retrieved results in the literature. WASP-39b and HAT-P-26b are excellent candidates for such an exercise as their transmission spectra extend all the way from the optical (\textit{STIS}) to near-infrared (\textit{Spitzer}). Their atmospheric properties have been inferred from ATMO Retrieval Code (ARC), which couples the ATMO models to a L-M least-squares minimizer and a Differential Evolution Chain Monte Carlo analysis \citep{tremblin_2015, wakeford_2017}. ATMO is used to compute 1-D T-P profiles for atmospheres in hydrostatic and radiative-convective equilibrium and calculate the transmission and emission spectra for a given atmospheric profile. We choose to compare with ARC because ATMO has been benchmarked against the Met Office SOCRATES code (used for the Earth) and because ATMO has a public scalable grid of transmission spectra \citep{goyal_2018b} that can be used for retrievals.  More importantly, the data for these planets provide relatively tight constraints on their atmospheric water abundance (WASP-39b) and heavy element abundance (HAT-P-26b) in the ATMO retrieval \citep{wakeford_2017, wakeford_2018}. These features provide useful tests for PLATON's retrieval capabilities and allow direct comparison of well constrained quantities using different retrieval models.

\begin{figure*}[ht]
  \centering \subfigure[HAT-P-26b] {\includegraphics
    [width=0.45\textwidth]{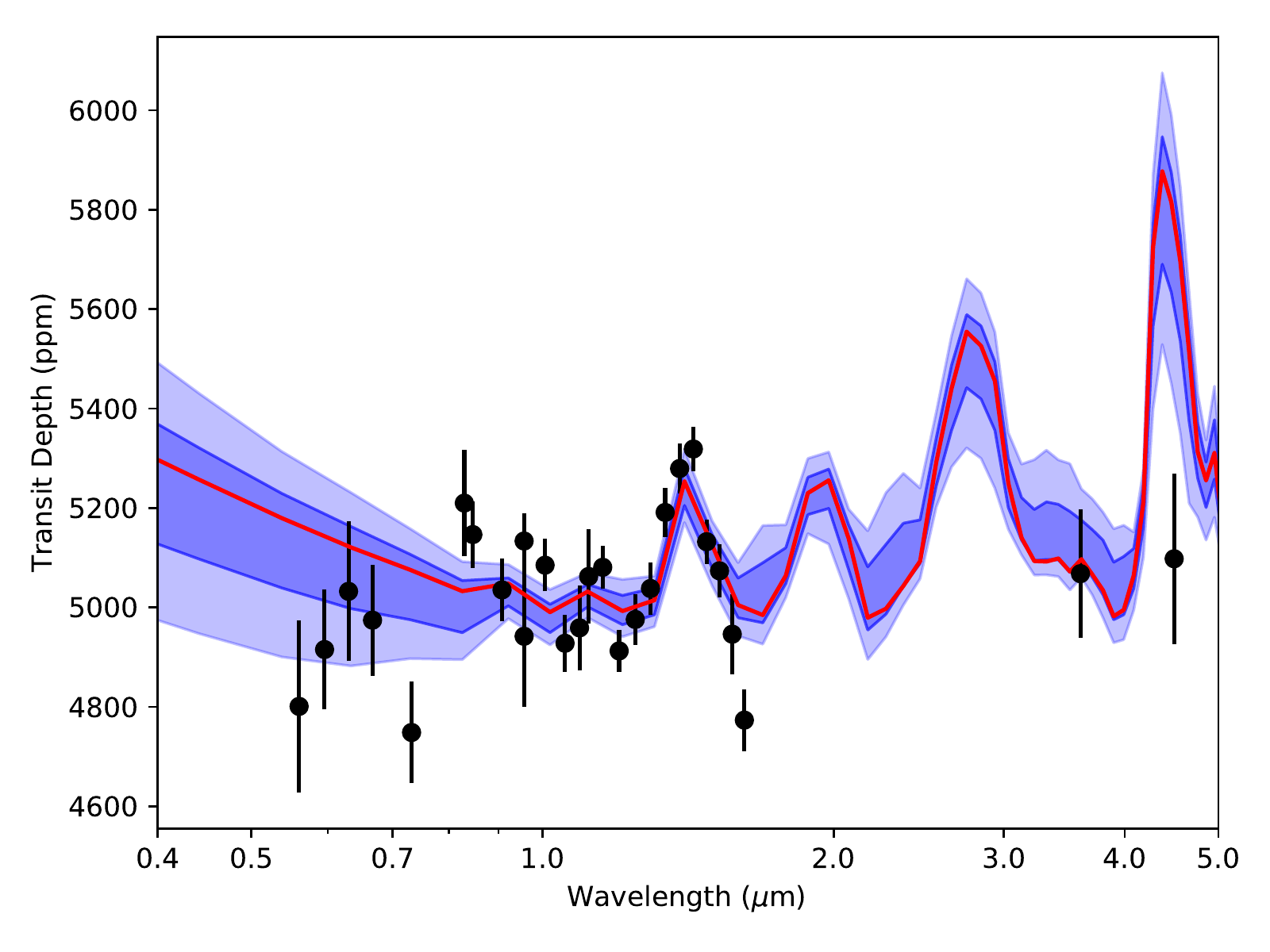}}\qquad
  \subfigure[WASP-39b] {\includegraphics
    [width=0.45\textwidth]{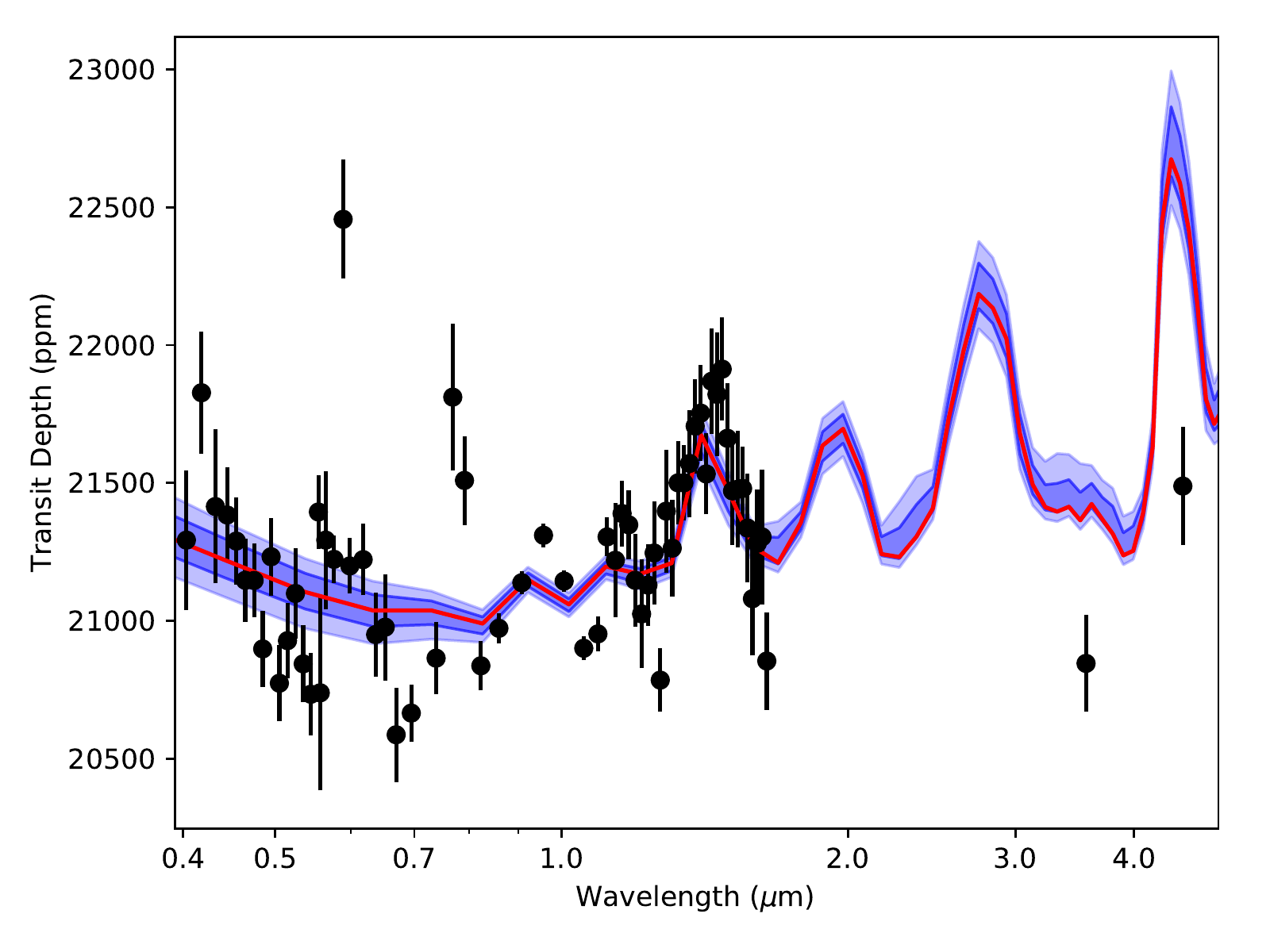}}
    \caption{The best fit transmission spectrum for HAT-P-26b and WASP-39b (red) and the $\pm 1 \sigma$ as well as the $\pm 2 \sigma$ uncertainties shown in blue.}
\label{fig:HAT26_WASP39_transmission}
\end{figure*}

\begin{figure*}[ht]
  \centering 
  \includegraphics[width=\textwidth]{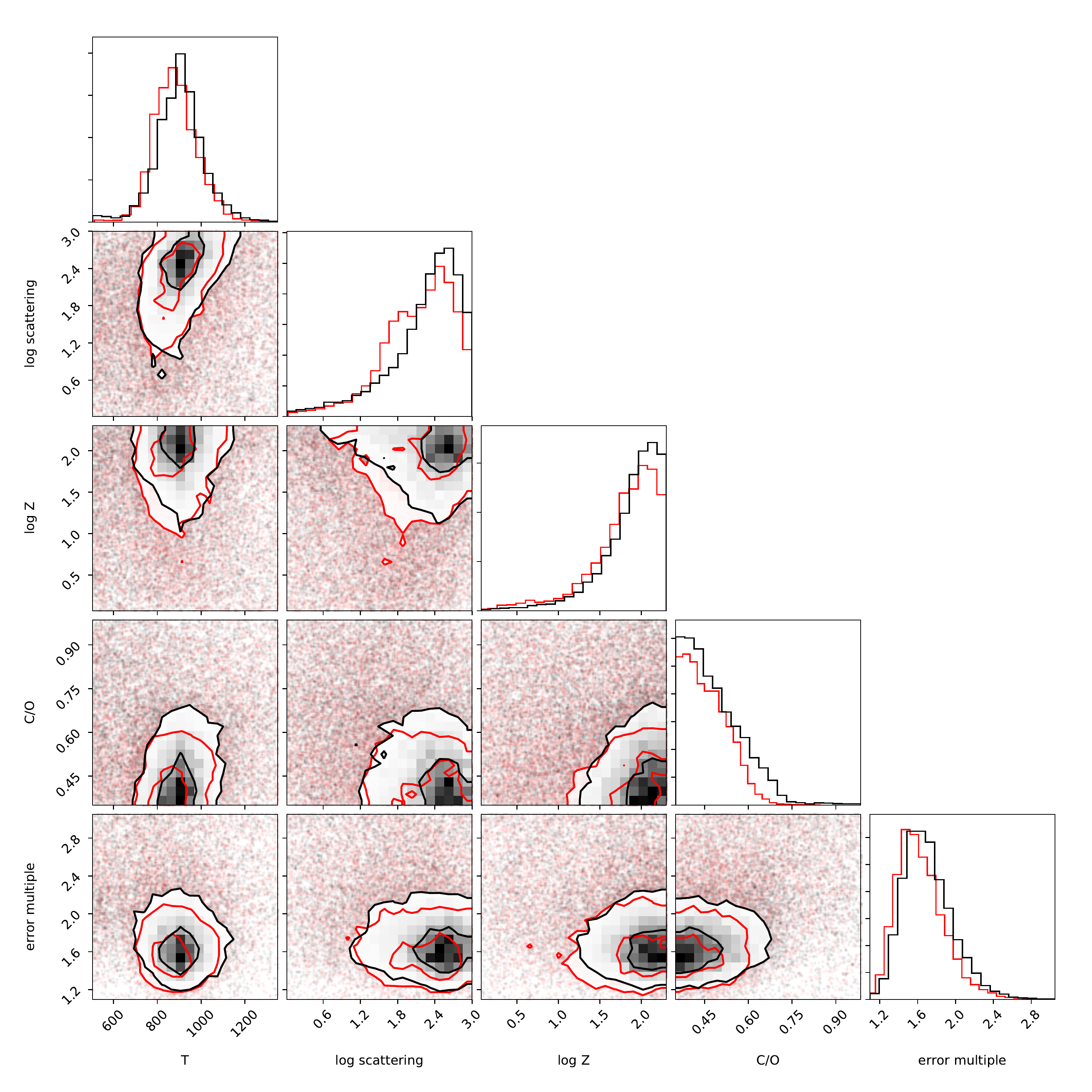}
  \caption{Posterior probability distribution of planetary and atmospheric properties for HAT-P-26b.  Red: retrieved with PLATON; black: retrieved with ATMO generic grid.}
\label{fig:HAT_P_26b_posteriors}
\end{figure*}

\begin{figure*}[ht]
  \centering 
  \includegraphics[width=\textwidth]{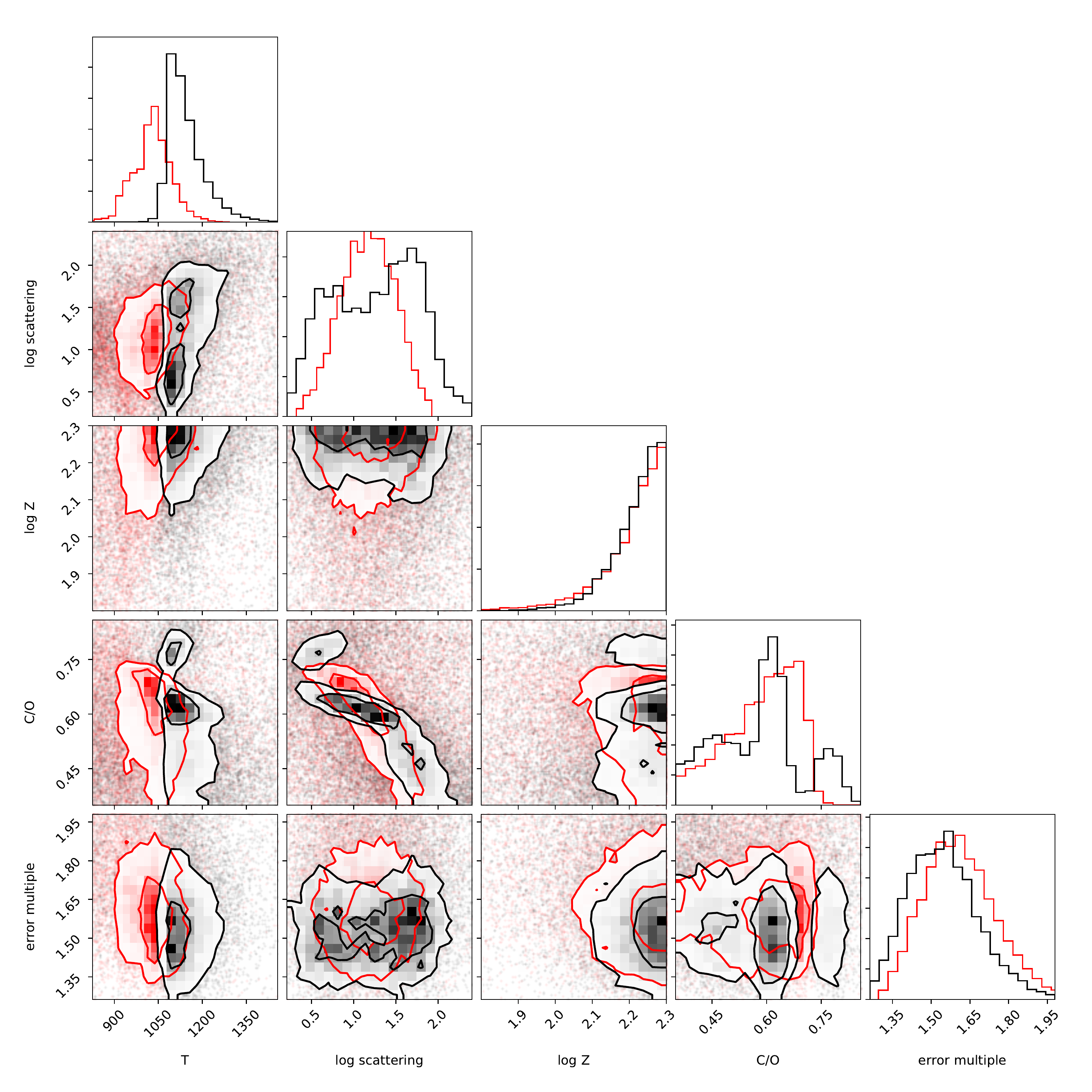}
  \caption{Posterior probability distribution of planetary and atmospheric properties for WASP-39b.  Red: retrieved with PLATON; black: retrieved with ATMO generic grid.}
\label{fig:WASP_39b_posteriors}
\end{figure*}

We run retrievals on these two planets with both PLATON and the ATMO grid.  We use the ATMO grid which accounts for rainout, which has 5 dimensions: temperature, gravity, metallicity, C/O ratio, scattering factor, and cloud parameter (representing a constant additional opacity across the atmosphere at all wavelengths).  Since there is no parameter in PLATON that corresponds exactly to the cloud parameter, we set the cloud parameter to zero in the ATMO models and the cloudtop pressure to infinity in the PLATON models.  This restricts our retrievals to cases without an optically thick cloud layer while still allowing for enhanced scattering due to optically thin hazes.  We use uniform priors for temperature, log(Z), C/O ratio, and log(scattering factor), with the minimum and maximum set to the minimum and maximum of the ATMO grid.  5D linear interpolation is used to compute the transmission spectrum for a certain set of parameters from the transmission spectra at the grid points, then scaled to the planetary and stellar radii.  For consistency, we use the same priors for PLATON, even though PLATON supports a wider parameter space.  
Figures \ref{fig:HAT_P_26b_posteriors} and \ref{fig:WASP_39b_posteriors} show the resulting posterior probability distributions for HAT-P-26b and WASP-39b, respectively, while the best fits are shown in Figure \ref{fig:HAT26_WASP39_transmission}.  HAT-P-26b shows strikingly good agreement between PLATON and ATMO, with all posteriors being nearly identical.  For WASP-39b, on the other hand, there are slight discrepancies.  PLATON favors a temperature $\sim$100 K lower, a discrepancy of roughly 1$\sigma$.  In addition, ATMO exhibits an odd multimodal distribution in scattering factor and C/O ratio that is not seen in PLATON.  This multimodal distribution is not seen in the retrieval from \cite{wakeford_2018}, which uses the ATMO code directly instead of using a generic grid.  The temperature retrieved by PLATON is also more consistent with \cite{wakeford_2018} than with the temperature we retrieved using the generic ATMO grid.  Additionally, we do not see a multimodal distribution if we use the ATMO grid which does not include rainout, nor do we see it if the observations of the two Na and K spectral features are excluded.  We therefore conclude that the odd behavior of our ATMO retrievals using the generic grid is likely due to a rare coincidence of factors.

\begin{table*}[t]
  \centering
  \caption{Best Fit Parameters for HAT-P-26b}
  \begin{tabular}{c C C C}
  \hline
  	Property & \text{PLATON} & \text{ATMO grid} & \text{\cite{wakeford_2017}}\\
   \hline
    Temperature (K) 				& 903 \pm 96  	 &  878^{+101}_{-87}\\
    C/O ratio						& 0.46^{+0.12}_{-0.08} &  0.45^{+0.09}_{-0.07}\\
    log$_{10}$(Z / Z$_{\odot}$) 	& 1.96^{+0.23}_{-0.39} & 1.87^{+0.27}_{-0.45} & 1.566 \pm 1.7034 \\
    log$_{10}$(Scattering Factor) 	& 2.37^{+0.38}_{-0.71} & 2.21^{+0.46}_{-0.62} \\
    error multiple                  & 1.68^{+0.29}_{-0.22} & 1.60^{+0.28}_{-0.21}\\
      \hline
  \end{tabular}
  \label{table:hat_p_26b_params}
\end{table*}

\begin{table*}[t]
  \centering
  \caption{Best Fit Parameters for WASP-39b}
  \begin{tabular}{c C C C}
  \hline
  	Property & \text{PLATON} & \text{ATMO grid} & \text{\cite{wakeford_2018}} \\
   \hline
    Temperature (K) 		 	 	&  1031^{+58}_{-69} & 1129^{+73}_{-38} & 1030^{+30}_{-20} \\
    C/O ratio						& 0.60^{+0.08}_{-0.12} & 0.59^{+0.1}_{-0.15} & 0.31^{+0.08}_{-0.05} \\
    log$_{10}$(Z / Z$_{\odot}$) 	& 2.23^{+0.05}_{-0.11} & 2.23^{+0.05}_{-0.08} & 2.18^{+0.12}_{-0.16} \\
    log$_{10}$(Scattering Factor) 	& 1.17^{+0.33}_{-0.34} & 1.30^{+0.49}_{-0.65}\\
      \hline
  \end{tabular}
  \vspace*{20px}
  \label{table:wasp_39b_params}
\end{table*}

Tables \ref{table:hat_p_26b_params} and \ref{table:wasp_39b_params} show the median and error of each retrieved parameter for HAT-P-26b and WASP-39b, respectively.  The error bars are derived by comparing the 84th and 16th percentiles to the median.  We have also included published retrieval results, when available.  For HAT-P-26b, all values are consistent.  For WASP-39b, PLATON favors lower temperature, but similar C/O ratio, metallicity, and scattering factor compared to the ATMO generic grid.  The published metallicity and temperature are both very similar to PLATON's results, while the published C/O ratio is lower than that of PLATON.  It is unclear why the published C/O ratio, which was retrieved by directly using the ATMO code, is lower than that retrieved by both PLATON and the ATMO grid.

\section{Precautions, Limitations, and Best Practices}
\begin{figure}[ht]
  \centering 
  \includegraphics[width=\linewidth]{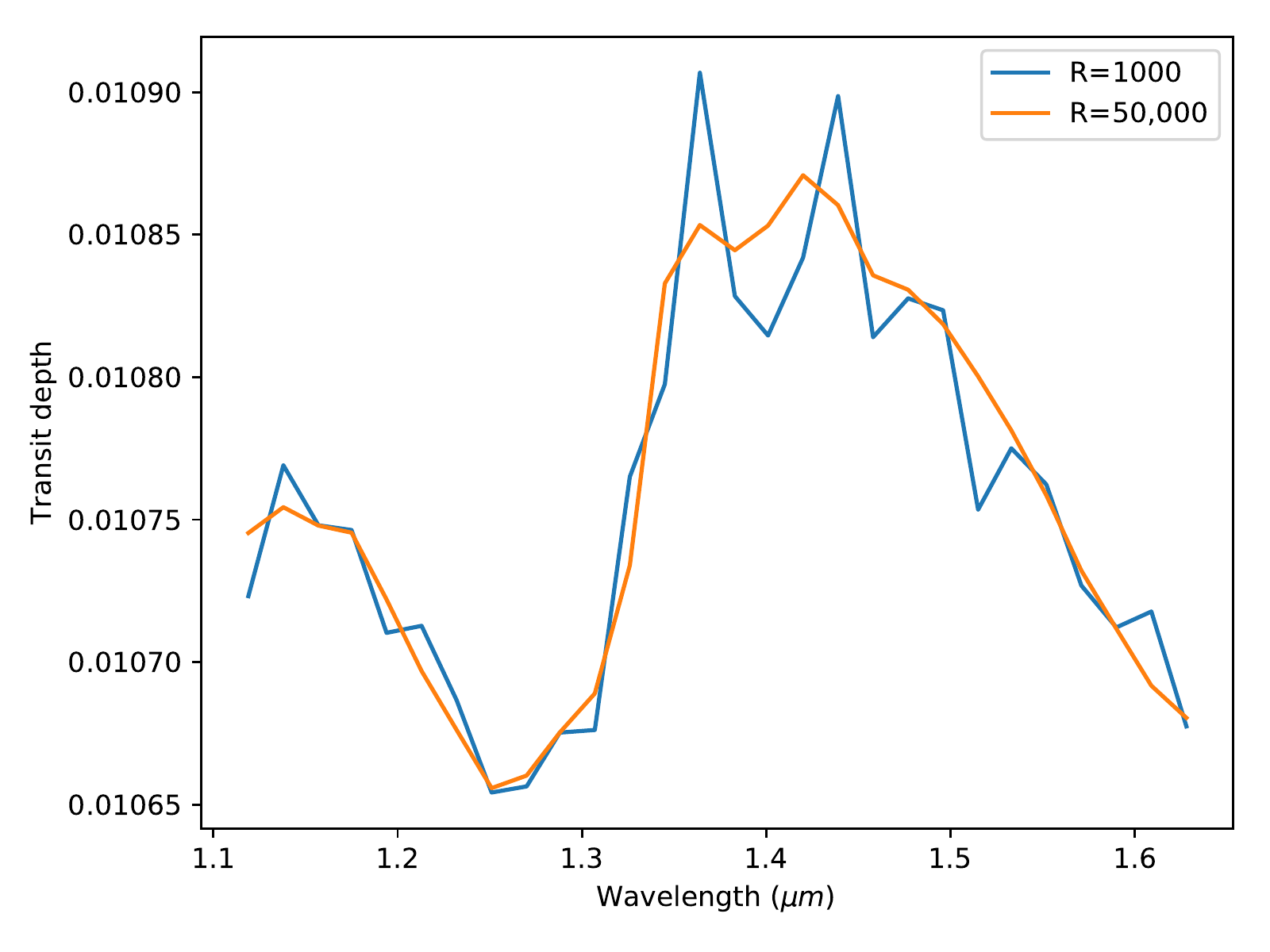}
  \caption{Transit depths across the WFC3 band for a 1200 K Jupiter orbiting a Sun-like star, computed assuming water-dominated opacity at two resolutions: R=1000 (native PLATON), and R=50,000 (assumed to be the truth).  The depths are binned with a bin size of 19 nm.  PLATON deviates from the true value by up to 50 ppm, although the deviation is much smaller at most wavelengths.}
\label{fig:resolution_comparison}
\end{figure}

PLATON is a powerful tool, but we were required to make some compromises in order to achieve our goal of a fast, easy to use package written in pure Python.  In this section we address some of the limitations of PLATON that users should be aware of.  The biggest source of error in PLATON comes from its relatively low R=1000 spectral resolution, which does not allow us to resolve individual lines at typical atmospheric pressures (P $<$ 1 bar).  When we sample these unresolved lines using our relatively coarse wavelength grid (this method is known as ``opacity sampling") it leads to spikiness in the cross sections and corresponding transmission/emission spectra, resulting in the errors visible in Figure \ref{fig:resolution_comparison}.  The idea behind opacity sampling is that even though the sampling resolution is much lower than that needed to resolve individual lines, it is still much higher than the instrumental resolution, and the spikiness in transit/eclipse depths can be smoothed out to an acceptable level by binning to instrumental resolution.

\begin{figure*}[ht]
  \centering 
  \includegraphics[width=\textwidth]{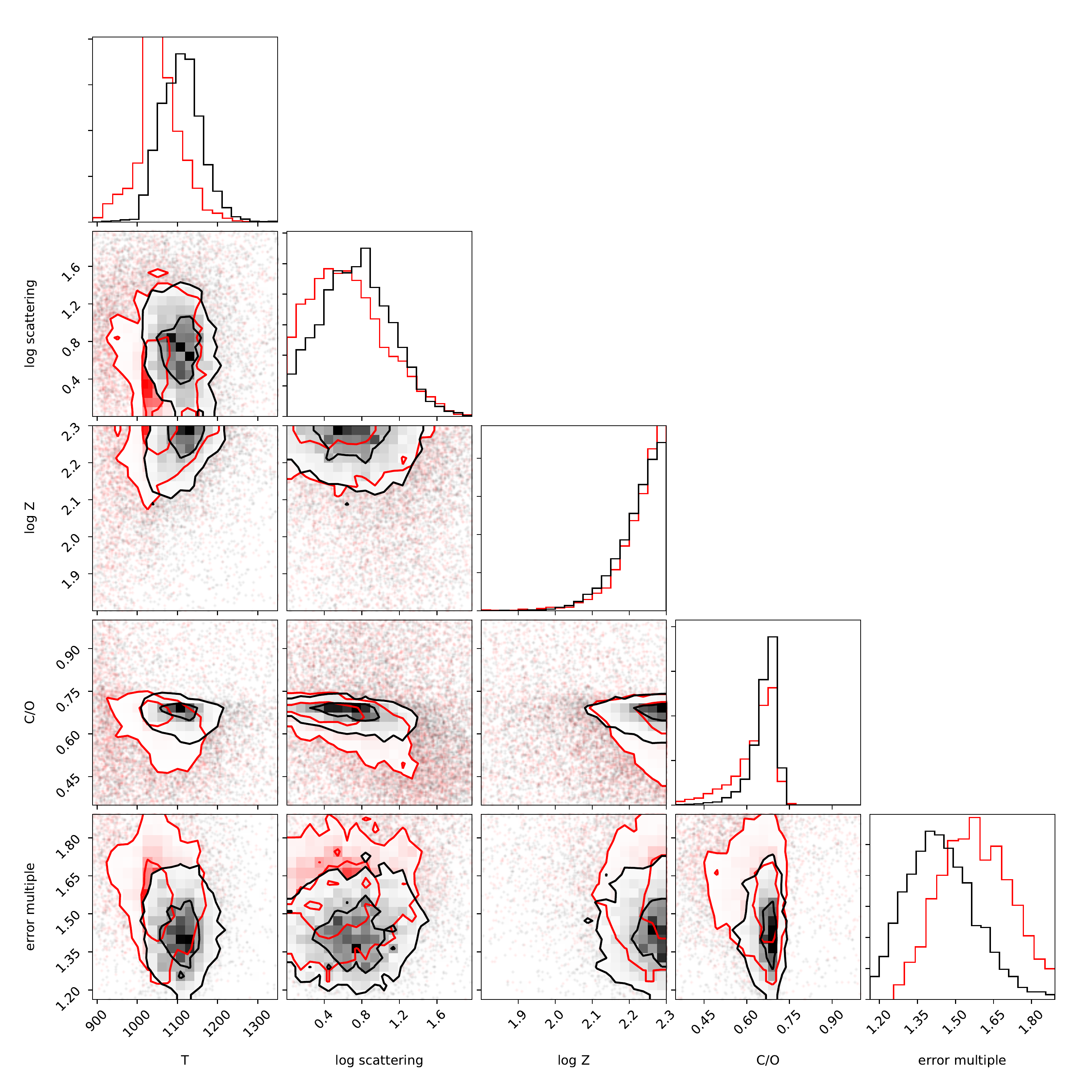}
  \caption{Comparison of retrievals on WASP-39b transit depths above 500 nm, using two opacity data resolutions: R=1000 (red) and R=10,000 (black).  As expected, the R=1000 retrieval has a higher error\_multiple and wider posteriors, reflecting the errors introduced by opacity sampling.  The low-resolution retrieval favors slightly lower temperatures (by 60 K), but the posteriors are otherwise very similar.}
\label{fig:compare_resolutions_corner}
\end{figure*}

In practice this means that users will obtain the best results when they utilize wavelength bins that are large relative to the intrinsic model resolution.  The number of wavelength grid points in a bin must be large enough to effectively average out the spikiness in transit depths caused by opacity sampling within that bin.  For example, a hot Jupiter at T=1200 K with solar metallicity has transit depths that fluctuate roughly 100 ppm from grid point to grid point at 1.4 $\mu$m, the middle of the HST WFC3 band.  At this wavelength, the wavelength grid points are separated by 1.4 nm, so a 30 nm wide wavelength bin would include 21 points.  The error generated by opacity sampling is then $\sim$100/sqrt(21) = 22 ppm, which is appreciably smaller than the typical published uncertainties on WFC3 data sets (i.e. \citealt{wakeford_2017, wakeford_2018}).  This sampling error is effectively ``white" in wavelength space-- i.e., it does not systematically bias the transit depth in one direction, but instead introduces random scatter around the true value.  If the observational data have errors comparable to or smaller than 22 ppm, retrievals with R=1000 will likely give only slightly wider posteriors than if we used line-by-line radiative transfer.  However, it is worth remembering that our model also makes other simplifying assumptions, such as an isothermal atmosphere, cloud opacity parameterization, and equilibrium chemistry, that may cause deviations from the measured transmission spectrum that are much greater than either the observational errors or the sampling errors.

If the default R=1000 opacities are too low resolution for a given retrieval, we also provide optional opacity files at R=2000 and R=10,000, available at \url{http://www.astro.caltech.edu/~mz/absorption.html}.  These only span 0.5-12 \um, not 0.3-30 \um, but the full wavelength coverage will be added in version 3.  Users can replace the existing opacity files in PLATON's data directory with a higher resolution version and then run PLATON as usual.  No code changes are necessary, but run time and memory usage will grow linearly with resolution.  

Our tests indicate that these high-resolution data are not usually necessary for STIS, WFC3, or Spitzer data. Figure \ref{fig:compare_resolutions_corner} compares a retrieval performed at R=10,000 to one performed at R=1000. The higher error\_multiple and wider posteriors caused by opacity sampling can clearly be seen, but the posteriors are otherwise very similar.  The high-resolution retrieval was 15 times slower, taking 215 minutes with 100 live points compared to 14 minutes at R=1000.  \cite{fisher_2018} performed a similar test--they tested resolutions of 1, 2, 5, and 10 cm\textsuperscript{-1} for WFC3 retrievals and found that ``for all of these values, the posterior distributions of T, $X_{H_2O}$, and $k_{\rm cloud}$ are somewhat similar."  The difference between the best and worst resolutions tested amounted to 0.5 sigma in temperature and water mixing ratio.  In addition, the 2D posterior distributions in their Figure 6 are very similar.  Our resolution of R=1000 corresponds to a wavenumber resolution of 6-9 cm\textsuperscript{-1}, so we are between their 5 and 10 cm\textsuperscript{-1} results in terms of the error introduced by opacity sampling. 

We next provide advice on how to ensure an accurate posterior when using nested sampling or MCMC.  For nested sampling, we have found that 100 live points is usually sufficient to obtain accurate 1D posteriors unless the posterior space is highly multimodal.  However, it is difficult to produce a publication-quality plot of the 2D posteriors with only 100 live points, as the low density of points in parameter space makes the contours look ragged and broken up.  We therefore recommend 100 live points for exploratory data analysis, and 1000 live points to generate publication-quality corner plots.  For MCMC, we find that 1000 steps with 50 walkers (resulting in 50,000 samples) is typically enough to sample more than 50 times the autocorrelation length.  We recommend 1000 steps for exploratory data analysis, and 10,000 steps to produce publication-quality posteriors.  For cases with widely separated multimodal distributions, we note that the MCMC method will perform poorly as walkers will have a hard time moving between peaks.  This is a well known limitation of MCMC, and we therefore recommend that users switch to nested sampling in these cases.

\subsection{Other Best Practices}
As its name implies, PLATON is designed to be accessible to users with minimal experience in modelling atmospheres.  To that end, we offer several suggestions to help newcomers do retrievals with minimal pain.  The first step in a retrieval is to decide which parameters to fix and which to retrieve, and following that, whether to use uniform or Gaussian priors.  The optimal choice varies on a case by case basis according to the quality of the user's data in comparison to the published data.  Nevertheless, there are a few guidelines that nearly always apply.  The stellar temperature, for example, has almost no impact on the transmission spectrum, and published values are always much more accurate than what one can derive from a retrieval on an exoplanet atmosphere.  The stellar temperature should be fixed.  Similarly, the spot fraction should be fixed to 0 for inactive stars, and restricted to an appropriately small range for active stars.  On the other hand, the metallicity and cloud-top pressure are typically not known in advance and should therefore be included as free parameters in the fit with log-uniform priors.  The stellar radius and planet mass usually have published values with an intermediate accuracy--not low enough that they should be ignored, but not high enough that they can be safely fixed to a single value.  This is a good use case for Gaussian priors, which take into account the published value's mean and standard deviation, while still allowing the parameter to vary within the published uncertainties as part of the fit.

We recommend that users begin with the parametric cloud and haze model and only switch to Mie scattering if the parametric model does not result in a good fit.  The parametric model is less physically motivated, but it has fewer free parameters and is less time consuming.  If the user does decide to try Mie scattering, we recommend leaving the particle size geometric standard deviation at the default value of 0.5.  This parameter has been measured for Earth aerosols (e.g. \citealt{shen_2015,pinnick_1978, elias_2009}) and ranges from 0.25 to 0.7, with a typical value of 0.5.  It is worth noting that the aerosol literature usually quotes $e^{\sigma_g}$ instead of $\sigma_g$.  When we refer to standard deviation, we always mean the $\sigma_g$ parameter in Equation \ref{eq:lognormal_dist}.

After choosing which parameters to fit, the user must pick default values and corresponding ranges for all model parameters used in the fit.  There is no point including a region of parameter space that is clearly unphysical, such as a temperature of 300 K for the most irradiated planet ever discovered, nor is there a point in restricting a parameter to a narrower range than justified by the current state of knowledge.

Next, the user might wonder whether to use MCMC or nested sampling.  In our experience it is best to start with nested sampling, because it is faster (usually finishing within minutes) and has a natural stopping point automatically determined by \nestle.  We have, however, found pathological cases where \nestle samples the parameter space extremely inefficiently; in these case, MCMC is necessary to have a reasonably fast retrieval.  With MCMC, it is necessary to set the number of walkers and the number of steps by hand, and checking for convergence is not trivial.  However, one advantage of MCMC is that it gives an unbiased sample from the posterior distribution, making it easy to plot the posterior probability distributions (also known as ``corner plots'').  We refer to \cite{foreman-mackey_2013} for tips on choosing \texttt{emcee} parameters.  With \nestle, the best one can do is get biased samples along with their weights, which can then be resampled into equal-weight samples.  Overall, the results from MCMC and nestled sampling are very similar, and we encourage users to try both to see which one they prefer.

\section{Conclusion}
We have developed a transmission spectrum calculator and retrieval tool in pure Python.  We release it on GitHub, and encourage the community to use it, contribute to it, and incorporate it in whole or part into other software.  This paper describes version 2.0 of the package, but we intend to keep it under continuous development--adding features, writing more unit tests, increasing user friendliness, and fixing bugs.  There is already a beta version of an eclipse depth calculator and retriever, and we intend to allow users to specify wavelength-dependent refractive indicies for Mie scattering in the near future.  

PLATON is not designed to model every physical phenomenon on exoplanets, or even to keep up with the cutting edge in theory.  Its niche is to be a fast, simple, and easily modifiable tool: a PLanetary Atmospheric Transmission tool for Observer Noobs.

\section{Acknowledgments}
M.Z. acknowledges Plato (Greek:$\Pi \Lambda\mathrm{AT}\Omega \mathrm{N}$) for his insightful philosophy and his contribution to the package's name. 

We thank Carlos E. Munoz for generating high-resolution opacity data in a format suitable for PLATON.  We thank Jayesh Goyal for assistance in comparing PLATON to ATMO.

Support for this work was provided by HST GO programs 13431, 13665, and 14260.  
\newline\newline
\textit{Software:} \texttt{numpy, scipy, matplotlib, emcee, nestle, corner, nose, Travis-CI}

\bibliographystyle{apj} \bibliography{main}

\end{document}